\documentclass[twocolumn,showpacs,preprintnumbers,amsmath,amssymb]{revtex4}
\usepackage{epsfig}
\usepackage{bbm}

\newcommand {\be}{\begin{equation}}
\newcommand {\ee}{\end{equation}}
\newcommand {\bea}{\begin{eqnarray}}
\newcommand {\ea}{\end{eqnarray*}}
\newcommand {\ba}{\begin{eqnarray*}}
\newcommand {\eea}{\end{eqnarray}}

\newcommand {\ham} {{\mathcal H}}

\newcommand {\bra}{\langle}
\newcommand {\ket}{\rangle}

\newcommand{\bm}[1]{ \mbox{\boldmath $#1$}  }

\begin{document}
\title{Integral relations and the adiabatic expansion method for 1+2 reactions above the 
breakup threshold: Helium trimers with soft-core potentials}

\author{E. Garrido}
\affiliation{Instituto de Estructura de la Materia, CSIC, Serrano 123, E-28006 Madrid, Spain}
\author{C. Romero-Redondo}
\affiliation{TRIUMF, 4004 Wesbrook Mall, Vancouver, BC, V6T 2A3, Canada}
\author{A. Kievsky and M. Viviani}
\affiliation{Istituto Nazionale di Fisica Nucleare, Largo Pontecorvo 3, 56100 Pisa, Italy}

\begin{abstract}
The integral relations formalism introduced in \cite{bar09,rom11}, and designed to describe 1+$N$ reactions, 
is extended here to collision energies above the threshold for the target breakup. These 
two relations are completely general, and in this work they are used together with the adiabatic 
expansion method for the description of 1+2 reactions.
The neutron-deuteron breakup, for which benchmark calculations are available, is taken as a test
of the method. The $s$-wave collision between the $^4$He atom and $^4$He$_2$ dimer above the breakup threshold and the possibility
of using soft-core two-body potentials plus a short-range three-body force will be investigated.
Comparisons to previous calculations for the three-body recombination and collision dissociation
rates will be shown.
\end{abstract}


\maketitle

\section{Introduction}

Calculation of continuum states corresponding to processes where a particle hits a bound $N$-body system 
requires in principle knowledge of the corresponding $(1+N)$-body wave function at large distances.
Needless to say, the technical difficulties one has to face in order to obtain the wave function increase
dramatically with $N$. In fact, already for $N=2$, calculation of the three-body wave function is far of
being trivial.

However, even if knowledge of the $(1+N)$-body wave function is unavoidable, the possibility of reducing
the distance at which such wave function is needed is in itself an important step forward in the 
description of the reaction. As shown in \cite{bar09,rom11}, this can indeed be done by means
of two integral relations that are based on the Kohn Variational Principle. 
These two integral relations are a generalization to more than two particles of the integral relation 
given in \cite{har67,hol72}, and they permit to obtain the ${\cal K}$- (or ${\cal S}$)-matrix of the 
reaction by using only the internal part of the wave function. Therefore, all the physical information 
concerning a given $1+N$ reaction can be obtained without an accurate knowledge of the asymptotic part
of the wave function.

The fact that the asymptotic part is not needed
anymore leads to a drastic reduction of the computer effort required to extract the ${\cal K}$-matrix. 
The only condition necessary to obtain accurate second-order estimates of the 
${\cal K}$-matrix through the integral relations is that the trial wave function must fulfill the 
Schr\"{o}dinger equation only in the interaction region. This means that, for instance, scattering states 
can be described with bound-state-like trial wave functions \cite{kie10}.

In \cite{bar09,rom11} the integral relations were implemented to describe $1+2$ reactions below the
breakup threshold. When used in combination with the hyperspherical adiabatic expansion method
the dimension of the ${\cal K}$-matrix describing the reaction is dictated by the number of adiabatic
potentials related to the possible outgoing elastic or inelastic channels, which typically is very
small. Furthermore, thanks to integral relations, the convergence of the ${\cal K}$-matrix in terms 
of the adiabatic channels included in the expansion of the wave function is highly accelerated. 
Actually, the pattern of convergence is similar to the one found when the same method is used for 
the description of bound states \cite{bar09,rom11}. This is in fact not a minor issue, 
since as proved in \cite{bar09b}, when used to describe low-energy scattering states, the convergence
of the adiabatic expansion slows down significantly, even to the point that an accurate calculation 
of the ${\cal K}$-matrix would require infinitely many adiabatic terms in the expansion, what in practice
makes the procedure useless.

The success of the method for energies below the breakup threshold immediately suggests its extension 
to describe low-energy breakup reactions. In this case the dimension of the ${\cal K}$-matrix is not finite, 
since contrary to the elastic and inelastic channels or transfer reactions, the full three-body continuum 
states are described by infinitely many adiabatic potentials.

The first goal of this work is to generalize the method described in \cite{bar09,rom11} for $1+N$ reactions 
to energies above the threshold for breakup of the bound target. This generalization will be shown in 
Section II, where a short summary of the method described in \cite{rom11} will be given. In 
Section III, the $n-d$ reaction, for which a series of benchmark calculations are available \cite{fri90,fri95}, 
will be used as a test of the method. 

The second goal, which will be discussed in Section IV, concerns the use of the new method to investigate
the $^4$He+$^4$He$_2$ atomic reaction above the dimer breakup threshold. The case of the $0^+$ state will 
be considered. In particular, we will focus on the use of soft-core $^4$He-$^4$He potentials.  
As shown in \cite{kie11}, the use of an attractive
gaussian potential reproducing the same two-body properties as a standard hard-core potential (like
for instance the LM2M2 potential) leads to equivalent bound three-body systems and phase-shifts for the elastic
$^4$He+$^4$He$_2$ reaction, but only once that the soft-core potential is used together with a three-body 
short-term force. The possibility of using the same kind of potentials also to describe the breakup 
channel is interesting, since it automatically eliminates all the technical difficulties arising from 
the presence of a hard-core repulsion in the potential (see for instance Ref.\cite{pen03}). 

Finally, we close this work with a short summary and the conclusions.

\section{Formalism}

In Refs.\cite{bar09,rom11} $1+2$ reactions were described within the framework of the hyperspherical adiabatic
expansion method for energies below the threshold for breakup of the dimer. Therefore, only elastic, inelastic,
and transfer processes were possible. Since the formalism is described in great detail
in Ref.\cite{rom11}, here we just summarize its main aspects, which are given in the first part
of this section. In particular, we  will focus on those key points that will permit an easier understanding of 
the generalization of the method to energies above the two-body breakup threshold, which will be shown in the 
second part of the section. In the last part we will describe the integral relations,
which are actually the tools that permit to extract the ${\cal K}$-matrix of the reaction from the
internal part of the wave functions.  

\subsection{Sketch of the method for energies below the breakup threshold}
\label{sub2a}

Given a three-body system, the corresponding wave function within the frame of the adiabatic
expansion method is written as:
\begin{equation}
\Psi(\bm{x},\bm{y})=\frac{1}{\rho^5} \sum_{n=1}^\infty f_n(\rho) \Phi_n(\rho,\Omega),
\label{eq1}
\end{equation}
where $\bm{x}$ and $\bm{y}$ are the usual Jacobi coordinates, and  $\{\rho,\Omega\}$ are the 
hyperradius and the five hyperangles obtained from $\bm{x}$ and $\bm{y}$ \cite{nie01}. The
wave function has a well defined total angular momentum $J$, but for simplicity in the notation 
we omit this index (and its projection) in the expression above.

In hyperspherical coordinates the Hamiltonian operator $\hat{\ham}$ takes the form:
\begin{equation}
\hat{\ham} =  -\frac{\hbar^2}{2 m} \hat{T}_\rho + \hat {\cal H}_\Omega    ,
\end{equation} 
where $\hat{T}_\rho$ is the hyperradial kinetic energy operator, $\hat {\cal H}_\Omega$ contains
all the dependence on the hyperangles, and $m$ is an arbitrary normalization mass.

The angular functions $\Phi_n(\rho,\Omega)$, which form the complete basis used for the wave function
expansion (\ref{eq1}), are actually the eigenfunctions of 
$\hat {\cal H}_\Omega$,
\begin{equation}
\hat { \cal H}_\Omega \Phi_n(\rho,\Omega)=\frac{\hbar^2}{2 m} \frac{1}{\rho^2}
\lambda_n(\rho) \Phi_n(\rho,\Omega),
\label{eq3}
\end{equation} 
in such a way that the adiabatic effective potentials
\begin{equation}
V^{(n)}_{eff}(\rho)=\frac{\hbar^2}{2m}\left( \frac{\lambda_n(\rho)+\frac{15}{4}}{\rho^2}-Q_{nn}(\rho) \right)
\label{eq4}
\end{equation}
enter in the coupled set of radial equations that permit to obtain the radial functions $f_n(\rho)$,
\begin{eqnarray}
&&\left[ -\frac{d^2}{d\rho^2} +  \frac{2m}{\hbar^2}(V^{(n)}_{eff}(\rho)-E)
 \right] f_n(\rho) \nonumber \\
&&+ \sum_{n'\neq n} \left( -2 P_{n n'} \frac{d}{d\rho} - Q_{n n'} \right)f_{n'}(\rho)
= 0   \label{eq4b}
\end{eqnarray}
where
\begin{eqnarray}
Q_{n n'}(\rho)&=&\Big\bra \Phi_n(\rho,\Omega) \Big|\frac{\partial^2}{\partial \rho^2} \Big|
                           \Phi_{n^\prime}(\rho,\Omega) \Big\ket_\Omega,\\
P_{n n'}(\rho)&=&\Big\bra \Phi_n(\rho,\Omega) \Big|\frac{\partial}{\partial \rho} \Big|
                           \Phi_{n^\prime}(\rho,\Omega) \Big\ket_\Omega,
\end{eqnarray}
where the subscript $\Omega$ indicates integration over the hyperangles only (see \cite{nie01} for details).

\begin{figure}
\epsfig{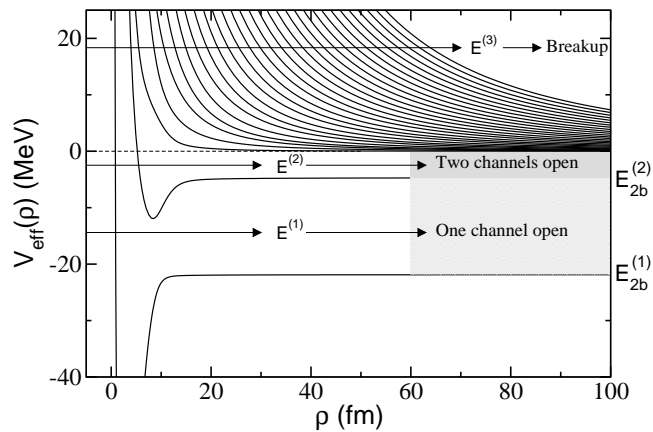}
\caption{ Typical effective adiabatic potentials for a three-body system where two two-body bound
states are present. The two lowest adiabatic
potentials go asymptotically to the binding energies $E_{2b}^{(1)}$ and $E_{2b}^{(2)}$ of the
two-body bound states. For a given three-body energy $E$, when $E_{2b}^{(1)}<E<E_{2b}^{(2)}$ only
one channel is open, when $E_{2b}^{(2)}<E<0$ both channels are open, and for $E>0$ the 
breakup channel is also open.}
\label{fig1}
\end{figure}

A typical behavior of the adiabatic potentials is shown in Fig.\ref{fig1}. They correspond
to a three-body system where two of the two-body subsystems have a bound state. This is reflected
in the fact that the two lowest effective adiabatic potentials go asymptotically to the binding
energies $E_{2b}^{(1)}$ and $E_{2b}^{(2)}$ of each bound two-body system.

In the figure the different regions defined by the energy of the incident particles are
depicted. All the three-body energies $E$ such that
$E_{2b}^{(1)} < E < E_{2b}^{(2)}$ (like $E^{(1)}$ in the figure) correspond to processes where
only one channel is open. Only the elastic collision between the third particle
and the bound two-body state with energy $E_{2b}^{(1)}$ is possible. When the three-body energy
increases up to the region $E_{2b}^{(2)} < E < 0$ ($E^{(2)}$ in the figure) a second channel is
open. Two different collisions are now possible, the one where a particle hits the bound state
with binding energy $E_{2b}^{(1)}$, and the one where a particle hits the state with binding
energy $E_{2b}^{(2)}$.
In the same way, each of these reactions has two possible outgoing channels, corresponding to
the two allowed bound two-body states and the third particle in the continuum. 
In other words, in this energy range the inelastic (if the two bound two-body states correspond 
to the same subsystem) or transfer (if the two bound two-body states correspond 
to different subsystems) channel is open.
When $E>0$ (like $E^{(3)}$ in the figure), the breakup channel is also open, and it is 
described by the remaining infinitely many adiabatic potentials.

The coupled system of radial equations (\ref{eq4b}) decouple asymptotically, and each of the radial wave functions behave
at large distances as dictated by:
\begin{equation}
\left(-\frac{\hbar^2}{2 m} \frac{d^2}{d\rho^2} +V_{eff}^{(n)}(\rho)-E \right)
 f_{n}(\rho)=0.
\label{eq5}
\end{equation}

When $n$ corresponds to a closed channel the radial wave function $f_n$ vanishes asymptotically. 
For values of $E$ below the two-body breakup threshold, as considered in \cite{rom11}, and
for a given incoming $1+2$ channel, only outgoing $1+2$ channels are allowed (either elastic,
inelastic, or transfer). In this case, as shown in \cite{nie01}, the equation above describing the
asymptotic behavior of the open-channel wave functions becomes
\begin{equation}
\left(\frac{d^2}{dy_n^2} +(k_y^{(n)})^2-\frac{\ell_y (\ell_y+1)}{y_n^2} \right)
 f_{n}(\rho)=0,
\label{eqas}
\end{equation}
where $\ell_y$ is the relative angular momentum between the dimer and the third particle,
$y_n$ refers to the modulus of the Jacobi coordinate between the center of mass of the outgoing bound 
two-body system and the third particle, and
\begin{equation}
k_y^{(n)}=\sqrt{ \frac{2m}{\hbar^2}(E-E_{2b}^{(n)}) },
\label{eq10} 
\end{equation}
with $E_{2b}^{(n)}$ being the binding energy of the bound two-body system associated to the
open channel $n$.

With this in hand, it is not difficult to see \cite{rom11} that 
the asymptotic form of the corresponding three-body wave function is given by:
\begin{equation}
\Psi_i \rightarrow
\sum_{n=1}^{n_0}
\left(
A_{in}^{(K)} F_{n}^{(K)} + B_{in}^{(K)} G_{n}^{(K)}
\right),
\label{eq7}
\end{equation}
where $i$ refers to the incoming channel, $n_0$ is the number of open channels (all of them $1+2$ channels), and
\begin{equation}
F_{n}^{(K)}=\sqrt{k_y^{(n)}} j_{\ell_y}(k_y^{(n)}y_n)\frac{1}{\rho^{3/2}} \Phi_n(\rho,\Omega) 
\label{eq8}
\end{equation}
\begin{equation}
G_{n}^{(K)}=\sqrt{k_y^{(n)}} \eta_{\ell_y}(k_y^{(n)}y_n)\frac{1}{\rho^{3/2}} \Phi_n(\rho,\Omega), 
\label{eq9}
\end{equation}
where $j_\ell$ and $\eta_\ell$ are the usual regular and irregular Bessel functions
(provided that we are dealing with short range potentials). 

As shown in \cite{rom11}, Eq.(\ref{eq7}) can be written in a compact matrix form as:
\begin{equation}
\Psi \rightarrow A^{(K)} F^{(K)} + B^{(K)} G^{(K)}=
A^{(K)} \left(F^{(K)} - {\cal K} G^{(K)}\right),
\label{eq11}
\end{equation}
where $\Psi$ is a column vector with $n_0$ terms corresponding to the $n_0$ open channels, $A^{(K)}$ and
$B^{(K)}$ are $n_0 \times n_0$ matrices made by the $A_{in}^{(K)}$ and $B_{in}^{(K)}$ elements in 
Eq.(\ref{eq7}), and $F^{(K)}$ and $G^{(K)}$ are again column vectors whose $n_0$ terms are given by 
Eqs.(\ref{eq8}) and (\ref{eq9}). 

From 
the equation above it is clear that the ${\cal K}$-matrix of the reaction is given by:
\begin{equation}
{\cal K}= -{A^{(K)}}^{-1} B^{(K)}.
\end{equation}
This matrix is real, and from it one can easily obtain the ${\cal S}$-matrix as 
$(1+i{\cal K})(1-i{\cal K})^{-1}$.

\subsection{Generalization to energies above the breakup threshold}
\label{gener}
When the total three-body energy $E$ in a $1+2$ reaction is above the threshold for breakup of the
dimer, the first consequence is that infinitely many adiabatic channels are then open (see
Fig.~\ref{fig1} for $E=E^{(3)}$). Of course, still
a finite number of them correspond to elastic, inelastic, or transfer processes, and the infinitely
many remaining ones describe the breakup channel. Therefore, in this case the corresponding 
${\cal K}$- (or ${\cal S}$-) matrix has infinite dimension.

In any case, for the finite open channels corresponding to outgoing $1+2$ structures, the expressions given
in the previous subsection are still valid. This means that for these particular outgoing channels
the three-body wave function behaves asymptotically as given by Eq.(\ref{eq7}), and Eqs.(\ref{eq8}) and 
(\ref{eq9}) are still valid.

On the other hand, the breakup channels are characterized by the fact that the effective potentials 
$V_{eff}^{(n)}$ associated to them go asymptotically to zero as:
\begin{equation}
V^{(n)}_{eff}(\rho) \stackrel{\rho \rightarrow \infty}{\rightarrow}
\frac{\hbar^2}{2m} \frac{\left( K+\frac{3}{2} \right) \left( K+\frac{5}{2} \right)}{\rho^2}, 
\label{eq13}
\end{equation}
where $K$ is the grand-angular quantum number defined as $2\nu+\ell_x+\ell_y$, where $\ell_x$ and $\ell_y$
are the orbital angular momenta associated to the Jacobi coordinates $\bm{x}$ and $\bm{y}$, 
respectively, and $\nu=0,1,2,\cdots$. Therefore, asymptotically, each breakup adiabatic potential is associated
to a fixed value of $K$. In fact, the corresponding angular eigenfunction $\Phi_n(\rho,\Omega)$ is,
also  asymptotically, a linear combination of hyperspherical harmonics with that particular
value of $K$.

When inserting (\ref{eq13}) into (\ref{eq5}), we easily obtain that, asymptotically, the radial wave 
function for an outgoing breakup channel $n$ satisfies the equation:
\begin{equation}
\left[\frac{d^2}{d\rho^2}+\kappa^2-\frac{(K+\frac{3}{2})(K+\frac{5}{2})}{\rho^2}
\right] f_{n}(\rho)=0,
\end{equation}
which is formally identical to the Eq.(\ref{eqas}), which is satisfied by the radial wave functions 
associated to outgoing $1+2$ channels, but replacing $\ell_y$ by $K+3/2$, 
$k_y^{(n)}$ by $\kappa=\sqrt{2mE/\hbar^2}$, and $y_n$ by $\rho$.

Therefore, in the more general case where the breakup channel is open, and assuming an incoming $1+2$ 
channel $i$, the asymptotic form of the corresponding three-body wave function is given by:
\begin{equation}
\Psi_i \rightarrow 
\sum_{n=1}^{\infty}
\left(
A_{in}^{(K)} F_{n}^{(K)} + B_{in}^{(K)} G_{n}^{(K)}
\right),
\label{eq15}
\end{equation}
where $F_n^{(K)}$ and $G_n^{(K)}$ are still given by Eqs.~(\ref{eq8}) and (\ref{eq9}), but of course, with the
understanding that when $n$ corresponds to an outgoing breakup channel the replacements
$\ell_y \rightarrow K+3/2$, $k_y^{(n)} \rightarrow \kappa$, and $y_n \rightarrow \rho$ have
to be made (note that the relations 
$j_{K+\frac{3}{2}}(z) = \sqrt{\frac{\pi}{2z}} J_{K+2}(z)$ and
$\eta_{K+\frac{3}{2}}(z) = \sqrt{\frac{\pi}{2z}} Y_{K+2}(z)$ permit
to write Eqs.(\ref{eq8}) and (\ref{eq9}) in terms of the Bessel functions $J_{K+2}$ and $Y_{K+2}$, which
is how the asymptotic form of the breakup channels is usually presented in the literature).

Of course, the matrix form in Eq.(\ref{eq11}) of the asymptotic wave function can still be used. 
As before, the ${\cal K}$-matrix of the reaction is given by ${\cal K}=-{A^{(K)}}^{-1} B^{(K)}$, 
but now the matrices $A^{(K)}$ and $B^{(K)}$ have in principle infinite dimension and some
truncation is then required. 

Due to the fact that the hyperradius $\rho$ appears as the natural 
radial coordinate describing the asymptotic behavior of the
breakup channels, we then have that, for these channels, the adiabatic 
expansion is able to reach the correct asymptotic behavior 
for a sufficiently large, but finite, value of $\rho$.
For this reason, for processes without $1+2$ open channels
($3 \rightarrow 3$ processes), the ${\cal K}$-matrix could 
in principle be extracted with sufficient accuracy from the asymptotic 
behavior. However, this is not true when $1+2$ channels are open. For these channels
the correct asymptotic form (described by (12) and (13)) can not be reached 
until the Jacobi coordinate $y$ and hyperradius $\rho$ are equal, which only 
happens at infinity. Therefore it is essential in this case to use a 
formalism in which the ${\cal K}$-matrix is not extracted from the asymptotic
part of the wave function but from its internal part.

\subsection{Integral relations}

The derivation of the integral relations has been shown in Ref.\cite{rom11}. This derivation is completely
general, and it is not particularized to the case of incident energies below the breakup threshold. 
Therefore, the same expressions derived in \cite{rom11} apply when the breakup channel is open.
According to it, by making use of the Kohn Variational Principle, it has been proved that when using a 
trial three-body wave function $\Psi^t$, one can obtain the matrices $A^{(K)}$ and $B^{(K)}$ 
accurate up to second order in $\delta(\Psi-\Psi^t)$, and their matrix elements are given by:
\begin{eqnarray}
        B_{ij} & = &  \frac{2m}{\hbar^2} 
   \langle\Psi_i^t | \hat{\cal H}-E |F_j^{(K)} \rangle \label{eq16} \\
        A_{ij} & = & - \frac{2m}{\hbar^2}  
  \langle \Psi_i^t|\hat{\cal H}-E |G_j^{(K)} \rangle \label{eq17},  
\end{eqnarray}
where $\Psi_i^t$ describes each possible incoming channel and the index $j$ refers to each
possible outgoing channel (either $1+2$ or breakup).

In this work, the trial three-body wave function will be the one obtained 
as sketched in subsection \ref{sub2a}. To be precise, it will be obtained by solving
the Faddeev equations by means of the hyperspherical adiabatic expansion method (see 
Ref.\cite{nie01} for details).

Since the regular and irregular functions $F^{(K)}$ and $G^{(K)}$ are
asymptotically solutions of $({\cal H}-E)F^{(K)},G^{(K)}=0$, it is then clear that
the integral relations (\ref{eq16}) and (\ref{eq17}) depend only on the
short-range structure of the scattering wave function $\Psi^t$. 

It is important to note that the function $G^{(K)}$, defined in Eq.(\ref{eq9}), is irregular at 
the origin. In order to avoid the problems arising from this fact, the $G^{(K)}$ function 
is regularized. This means that in Eq.(\ref{eq17}) the Bessel function
$\eta$ contained in $G^{(K)}$ is actually replaced by another function that goes to zero at the origin 
and behaves exactly as $\eta$ at large distances. In \cite{rom11} this was done by using
$\tilde{\eta}_\ell(z)=(1-e^{-\gamma z})^{\ell+1} \eta_\ell(z)$, where $\gamma$ ($>0$) is a 
parameter. However, in our case, where the index $\ell$ can reach pretty high values 
($\ell=K+3/2$ in the breakup channels) this procedure is not appropriate. 

In this work we have regularized the irregular Bessel function
by solving Eq.(\ref{eq5}) for each individual adiabatic potential. The solutions of this
equation behave asymptotically as $f_n(z) \rightarrow z j_\nu(z)-\tan \delta_n z \eta_\nu(z)$, 
where $\delta_n$ is the phase-shift, and the index $\nu$ is equal to $\ell_y$ for the $1+2$ 
channels, and $K+3/2$ for the breakup channels. Then, the irregular Bessel function 
implicitly contained in Eq.(\ref{eq17}) is replaced by:
\begin{equation}
\tilde{\eta}_\nu(z)=\frac{z j_\nu(z)-f_n(z)}{z \tan \delta_n},
\end{equation}
which by construction is regular at the origin and goes asymptotically to $\eta_\nu(z)$.

\section{A test case: neutron-deuteron scattering}

The benchmark solutions for neutron-deuteron breakup amplitudes are shown in \cite{fri90,fri95}.
For this reason we take this case as a test for the method shown in this work.

In particular, the nucleon-nucleon interaction is chosen to be the revised Malfliet-Tjon I-III model $s$-wave
potential \cite{fri90}, which for the spin triplet and singlet cases takes the form:
\begin{eqnarray}
V_t(r)&=&\frac{1}{r}\left( -626.885 e^{-1.55 r} +1438.72 e^{-3.11 r} \right) \label{eq19}\\
V_s(r)&=&\frac{1}{r}\left( -513.968 e^{-1.55 r} +1438.72 e^{-3.11 r} \right) \label{eq20},
\end{eqnarray}
where $r$ is given in fm, and the potential in MeV. Also, $\hbar^2/m$=41.47 MeV fm$^2$. The potential 
above leads to a binding energy for the deuteron of 2.2307 MeV.

\begin{figure}
\epsfig{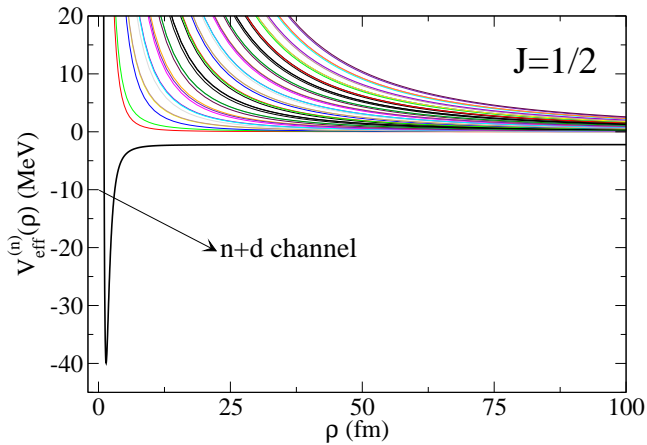}
\caption{(Color online) Effective adiabatic potentials for the neutron-neutron-proton system in the 
doublet case. The thick
curve corresponds to the neutron+deuteron channel, and it goes asymptotically to the deuteron binding
energy. All the remaining adiabatic potentials correspond to breakup channels.}
\label{fig2}
\end{figure}

In the calculation only $s$-waves are considered. Therefore, two different total angular momenta are possible, 
the quartet case ($J=3/2$), for which only the triplet $s$-wave potential (\ref{eq19}) enters, 
and the doublet case ($J=1/2$), for which both, the singlet and the triplet potentials contribute.
In Fig.\ref{fig2} we show the computed adiabatic effective potentials (Eq.(\ref{eq5})) 
for the doublet case. As you can see, they follow the general trend of the potentials shown in Fig.\ref{fig1},
although in this case there is only one $1+2$ channel, which corresponds to the neutron-deuteron reaction
that we want to investigate. The corresponding adiabatic potential is given by the thick-solid line in the 
figure, and it will be labeled as channel 1. As expected, the asymptotic value of this potential corresponds
to the binding energy of the deuteron.

\subsection{Inelasticity and phase-shifts}

The unitarity of the ${\cal S}$-matrix implies that given an incoming channel, for instance channel 1 
($n+d$ channel), we have that $\sum_1^\infty |{\cal S}_{1n}|^2=1$, or, in other words, 
\begin{equation}
\sum_{n=2}^\infty |{\cal S}_{1n}|^2=1-|{\cal S}_{11}|^2,
\end{equation}
which means that an accurate calculation of the elastic term ${\cal S}_{11}$ amounts to an accurate
calculation of the infinite summation of the $|{\cal S}_{1n}|^2$ terms ($n>1$) corresponding the 
breakup channels.

Also, the complex value of ${\cal S}_{11}$ can be written in terms of a complex phase-shift $\delta$ as:
\begin{equation}
{\cal S}_{11}=e^{2i\delta}=e^{-2 \mbox{\scriptsize Im}(\delta)} e^{2i \mbox{Re}(\delta)}
=|{\cal S}_{11}|e^{2i \mbox{Re}(\delta)}.
\label{eq22}
\end{equation}

\begin{table}
\caption{Inelasticity parameter $|{\cal S}_{11}|$ for the neutron-deuteron scattering for two
different laboratory neutron beam energies (14.1 MeV and 42.0 MeV) for the doublet and
quartet cases. The value of $K_{\mbox{\scriptsize max}}$ is the $K$-value associated 
to the last adiabatic potential included in the calculation. The numbers within parenthesis
have been obtained without use of the integral relations. 
The last row gives the value quoted in Ref.\cite{fri95}.}
\label{tab1}
\begin{tabular}{c|cc|cc}
  $K_{\mbox{\scriptsize max}}$   & \multicolumn{2}{c|}{Doublet} & \multicolumn{2}{c}{Quartet} \\ \hline
       &      14.1 MeV         &       42.0 MeV       &       14.1  MeV        &     42.0  MeV          \\ \hline
   4   &    {\bf 0.4662}       &     {\bf 0.4929}     &     {\bf 0.9794}       &   {\bf 0.8975}         \\
       &{\scriptsize (0.4710)} &{\scriptsize (0.4719)}&{\scriptsize (0.9809)}  & {\scriptsize (0.8865)} \\
   8   &    {\bf 0.4637}       &     {\bf 0.4993}     &     {\bf 0.9784}       &   {\bf 0.9026}         \\
       &{\scriptsize (0.4670)} &{\scriptsize (0.4985)}&{\scriptsize (0.9794)}  & {\scriptsize (0.9050)} \\
  12   &    {\bf 0.4640}       &     {\bf 0.5014}     &     {\bf 0.9783}       &   {\bf 0.9030}         \\
       &{\scriptsize (0.4664)} &{\scriptsize (0.5041)}&{\scriptsize (0.9792)}  & {\scriptsize (0.9071)} \\
  16   &    {\bf 0.4643}       &     {\bf 0.5019}     &     {\bf 0.9782}       &   {\bf 0.9031}         \\
       &{\scriptsize (0.4666)} &{\scriptsize (0.5051)}&{\scriptsize (0.9792)}  & {\scriptsize (0.9071)} \\
  20   &    {\bf 0.4644}       &     {\bf 0.5021}     &     {\bf 0.9782}       &   {\bf 0.9033}         \\
       &{\scriptsize (0.4666)} &{\scriptsize (0.5052)}&{\scriptsize (0.9791)}  & {\scriptsize (0.9069)} \\
  24   &    {\bf 0.4645}       &     {\bf 0.5022}     &     {\bf 0.9782}       &   {\bf 0.9033}         \\
       &{\scriptsize (0.4666)} &{\scriptsize (0.5055)}&{\scriptsize(0.9790)}   & {\scriptsize (0.9071)} \\
  28   &    {\bf 0.4645}       &     {\bf 0.5022}     &     {\bf 0.9782}       &   {\bf 0.9033}         \\
       &{\scriptsize (0.4667)} &{\scriptsize (0.5056)}&{\scriptsize(0.9790)}   & {\scriptsize (0.9071)} \\
                                                                                     \hline
Ref.\cite{fri95} &    {\bf 0.4649}    &  {\bf 0.5022}    &   {\bf 0.9782}      &   {\bf 0.9033}   \\ 
\end{tabular}
\end{table}

The value of $|{\cal S}_{11}|^2$ gives the probability of elastic neutron-deuteron scattering, and 
$|{\cal S}_{11}|$ is what usually referred to as the inelasticity parameter (denoted by $\eta$ in 
\cite{fri90,fri95}).  Obviously, the closer the inelasticity to 1 the more elastic the reaction.
In fact, for energies below the breakup threshold the phase-shift is real and $|{\cal S}_{11}|=1$.

In Table~\ref{tab1} we give the inelasticity parameter $|{\cal S}_{11}|$ for the two laboratory neutron
energies used in \cite{fri95}, i.e., 14.1 MeV and 42.0 MeV. We have computed the inelasticity
parameter for both, the doublet and the quartet states. We give the computed values of $|{\cal S}_{11}|$ 
for different truncations in the infinite summation (\ref{eq15}). In particular, in the table we
give the value $K_{\mbox{\scriptsize max}}$ of the asymptotic grand-angular quantum number associated to the last adiabatic 
potential included in the calculation (see Eq.(\ref{eq13})). The values given in the table without
parenthesis have been obtained by using the integral relations (\ref{eq16}) and (\ref{eq17}), while
the ones within parenthesis have been calculated from the $A^{(K)}$ and $B^{(K)}$ matrices extracted
directly from the asymptotic part of the three-body wave function, as indicated in Eq.(\ref{eq11}).
The last row in the table gives the value obtained in \cite{fri95}.

As seen in the table, when using the integral relations the agreement with the results in \cite{fri95} is 
very good. Actually, we obtain precisely the same result for the two energies in the quartet case,
and a tiny difference clearly smaller than 0.1\% in the doublet case. Furthermore, the pattern of 
convergence is rather fast, specially in the quartet case, for which already for
$K_{\mbox{\scriptsize max}}=8$ we obtain a result that can be considered very accurate. In the
doublet case the convergence is a bit slower, and a value of $K_{\mbox{\scriptsize max}}$ of about
16 is needed.  As we can see from the values within parenthesis, when the integral relations
are not used, the value of $|{\cal S}_{11}|$ seems to converge more slowly, and even if converged, the result
is less accurate. 

\begin{table}
\caption{The same as Table~\ref{tab1} for Re($\delta$).}
\label{tab2}
\begin{tabular}{c|cc|cc}
  $K_{\mbox{\scriptsize max}}$   & \multicolumn{2}{c|}{Doublet} & \multicolumn{2}{c}{Quartet} \\ \hline
        &         14.1 MeV     &      42.0 MeV      &      14.1  MeV      &    42.0  MeV         \\ \hline
   4    &      {\bf 105.82}    &    {\bf 42.66}     &    {\bf 69.04}      &   {\bf 38.98}        \\
        & {\scriptsize(97.62)} &{\scriptsize(28.89)}&{\scriptsize(60.88)} & \scriptsize{(25.29)} \\
   8    &      {\bf 105.57}    &    {\bf 41.65}     &    {\bf 68.99}      &   {\bf 37.95}        \\
        & {\scriptsize(99.91)} &{\scriptsize(32.88)}&{\scriptsize(63.24)} & \scriptsize{(28.98)} \\
  12    &      {\bf 105.53}    &    {\bf 41.49}     &    {\bf 68.98}      &   {\bf 37.77}        \\
        & {\scriptsize(101.01)}&{\scriptsize(34.80)}&{\scriptsize(64.43)} & \scriptsize{(31.07)} \\
  16    &      {\bf 105.53}    &    {\bf 41.46}     &    {\bf 68.97}      &   {\bf 37.73}        \\
        & {\scriptsize(101.68)}&{\scriptsize(36.05)}&{\scriptsize(65.11)} & \scriptsize{(32.28)} \\
  20    &      {\bf 105.53}    &    {\bf 41.45}     &    {\bf 68.96}      &   {\bf 37.72}        \\
        & {\scriptsize(102.12)}&{\scriptsize(36.81)}&{\scriptsize(65.54)} & \scriptsize{(33.03)} \\
  24    &      {\bf 105.53}    &    {\bf 41.44}     &    {\bf 68.96}      &   {\bf 37.71}        \\
        & {\scriptsize(102.41)}&{\scriptsize(37.30)}&{\scriptsize(65.84)} & \scriptsize{(33.55)} \\
  28    &      {\bf 105.53}    &    {\bf 41.44}     &    {\bf 68.96}      &   {\bf 37.71}        \\
        & {\scriptsize(102.49)}&{\scriptsize(37.90)}&{\scriptsize(65.88)} & \scriptsize{(33.58)} \\
                                                                                                   \hline
Ref.\cite{fri95}   &  {\bf 105.50}    &   {\bf 41.37}     &   {\bf 68.96}       &    {\bf 37.71}   \\ 
\end{tabular}
\end{table}

From Eq.(\ref{eq22}) we have that, together with the inelasticity parameter $|{\cal S}_{11}|$, a complete
specification of the matrix element ${\cal S}_{11}$ requires knowledge of the real part of the phase-shift 
Re($\delta$). The corresponding computed values are shown in Table~\ref{tab2}, where the meaning of
the different columns is the same as in Table~\ref{tab1}. The behavior of Re($\delta$) is similar to
what shown in Table~\ref{tab1} for $|{\cal S}_{11}|$. For the quartet case the same results as in \cite{fri95}
are obtained, while for the doublet again a very small difference smaller than 0.1\% is again found. 
The pattern
of convergence is also similar. A $K_{\mbox{\scriptsize max}}$ value of 12 is already enough to
get a quite accurate value. The main difference compared to the results for $|{\cal S}_{11}|$ shown 
in Table~\ref{tab1} is that now the values of Re($\delta$) obtained without use of the integral
relations are much less converged and much less accurate. The difference with the true result can
reach up to 10\%. This behavior was already observed in \cite{bar09} for the phase-shift
in a reaction below the breakup threshold. This is due to the fact that the hyperradius $\rho$ and the 
Jacobi coordinate $y$ entering in the asymptotic forms (\ref{eq8}) and (\ref{eq9}) are equivalent
only at infinity as commented at the end of section \ref{gener} . This means that a correct extraction of the phase-shift from the asymptotic part
of the wave function requires to impose the boundary condition at infinity, for which also infinite
adiabatic terms would be needed.

Therefore, for energies above the breakup threshold, the conclusion from Tables~\ref{tab1} and \ref{tab2} 
is similar to the one reached in \cite{bar09} for $1+2$ elastic processes, that is, the 
use of the integral relations is crucial from two different points of view. First, it accelerates drastically 
the convergence of the values of the ${\cal S}$-matrix elements, and second, they are needed
in order to obtain the correct result.  

%
%

\section{Soft-core $\mbox{$^4$He}$-$\mbox{$^4$He}$ potential and the $\mbox{$^4$He}$-$\mbox{$^4$He}_2$ reaction}

Once the integral relations have been proved to be efficient in order to describe
$1+2$ reactions above the breakup threshold, in this section we shall use them to study
the atomic $\mbox{$^4$He}$-$\mbox{$^4$He}_2$ process. In particular, we shall focus on three-body
states with spin and parity $0^+$, and the possibility of using simple two-body soft-core potentials
will be investigated.

The $\mbox{$^4$He}$-$\mbox{$^4$He}$ molecule is known to be one of the biggest diatomic molecules. Its 
binding energy has been
estimated to be around 1 mK with a scattering length $a$ around 190 a.u \cite{luo96,gri00}. 
Different accurate investigations of the $\mbox{$^4$He}$-$\mbox{$^4$He}$ interaction are available 
in the literature \cite{azi91,tan95,kor97,jez07}. All of them present the common feature of a sharp 
repulsion below an interparticle distance of approximately 5 a.u.. The presence of the repulsive core
is the source of a series of important technical difficulties. For instance, the wave function
in the inner regions, which is very small due to the large potential repulsion, is decisive for the 
energy of the bound states or the asymptotic properties of the continuum states. Therefore, the
wave function must be calculated with high accuracy in this region, which typically requires a very
important increase of the basis size. Furthermore, in case of using the adiabatic expansion method,
the angular eigenvalues $\lambda_n(\rho)$ (see Eq.(\ref{eq3})) also diverge for small $\rho$, and this
divergence provokes very frequent crossings between them that sometimes are not easy to handle \cite{nie98}.

In Ref.\cite{kie11} the possibility of using a soft-core potential was investigated in the context
of $^4$He-$\mbox{$^4$He}_2$ collisions below the threshold for breakup of the dimer. In particular
the gaussian potential suggested in \cite{nie98}
\begin{equation}
V_{2b}(r)=-1.227 e^{-r^2/10.03^2}
\label{soft}
\end{equation}
was used (the strength of the potential is in K and the range in a.u.).

\begin{table*}
\begin{center}
\caption{Helium dimer and helium trimer properties obtained with the 
soft-core potential in Eq.(\ref{soft}), the hard-core potential LM2M2 in Ref.\cite{azi91},
the modified soft-core potential (soft-core 2) in Eq.(\ref{soft2}), and the hard-core SAPT potential
in Ref.\cite{jez07}, respectively. The two-body properties are given in the upper part
of the table, where $E_{2b}$, $a$, $r_0$, and $\langle r \rangle$ are the dimer binding energy, the
scattering length, the effective range, and the average interatomic distance, respectively. The three-body
properties are given in the lower part, where $E_{3b}^{(g.s.)}$, $E_{3b}^{(exc.)}$, and
$a_0$ are the binding energy of the trimer ground state, the binding energy of the trimer excited state,
and the atom-dimer scattering length, respectively. For the soft-core and soft-core-2 potentials the
column corresponding to the trimer properties has been split in two. The results given in the
left part have been obtained without inclusion of any effective three-body force. The ones
given in the right part are obtained after inclusion of a three-body force fitted to match the
binding energy of the trimer ground state obtained with the LM2M2 \cite{kol09} and 
SAPT \cite{jez07,sun08} potentials, respectively.
The parameters used for the three-body potential are denoted by $W_0$ and $\rho_0$ (see Eq.(\ref{3b})).
All the distances are given in a.u. and the energies in mK, except $W_0$ which is given in K.}
\label{tab3}
\begin{ruledtabular}
\begin{tabular}{ccccccc}
           &  \multicolumn{2}{c}{Soft-core} & LM2M2 \cite{kol09}     
               & \multicolumn{2}{c}{Soft-core 2} &   SAPT \cite{jez07,sun08}   \\ \hline
 $E_{2b}$ (mK) &  \multicolumn{2}{c}{$-1.296$}  &  $-1.302$ & \multicolumn{2}{c}{$-1.554$}    &  $-1.564 $\\
 $a$  (a.u.) &  \multicolumn{2}{c}{189.95}    &  189.05   & \multicolumn{2}{c}{174.09}      &  173.50   \\
 $r_0$  (a.u.) &  \multicolumn{2}{c}{13.85}     &  13.84    & \multicolumn{2}{c}{13.80}       &  13.79    \\
 $\langle r \rangle$ (a.u.) & \multicolumn{2}{c}{98.4}  & 98.2 &   \multicolumn{2}{c}{90.5}   &   90.3 \\ \hline
 ($W_0$, $\rho_0$) (K, a.u.)&   $(0,-)$   &  $(18.314,6)$ &          &   $(0,-)$  &  $(17.760,6)$ &     \\
 $E_{3b}^{(g.s.)}$ (mK)  &   $-150.0$     &   $-126.4$  &  $-126.4$  &   $-154.9$ &   $-130.9$    &  $-130.9$ \\
 $E_{3b}^{(exc.)}$ (mK)  &   $-2.467$     &   $-2.287$  &  $-2.265$  &   $-2.805$ &   $-2.612$    &  $-2.588$ \\
 $a_0$     (a.u.)        &   165.9        &   210.6     &  224.3     &     181.7  &   226.0         &  226.8 \\
\end{tabular}
\end{ruledtabular}
\end{center}
\end{table*}

The dimer properties obtained with this potential are given in the second column in the upper part 
of table~\ref{tab3}. In particular, the dimer binding energy ($E_{2b}$), the two-body scattering 
length ($a$), the effective range ($r_0$), and the interatomic distance ($\langle r \rangle$) are given. 
The two parameters in the soft-core gaussian potential (\ref{soft}) were fitted to reproduce the scattering 
length and the effective range of the hard-core potential LM2M2 \cite{azi91}. As seen in the third column of 
Table~\ref{tab3} (upper part), when this is done the binding energy $E_{2b}$ and the 
interatomic distance $\langle r \rangle$ are also well reproduced.

However, even if both potentials have the same two-body properties, when moving to the three-body states 
important differences appear. The soft-core potential overbinds the two bound states in $\mbox{$^4$He}_3$ and 
the atom-dimer scattering length is clearly smaller (second and fourth columns in the lower part of 
Table~\ref{tab3}). In fact, as shown in \cite{kie11}, the phase-shifts obtained with these two 
potentials for the $\mbox{$^4$He}-\mbox{$^4$He}_2$ elastic scattering
(below the breakup threshold) clearly differ from each other. As an example, for an incident energy
of 1 mK, the phase-shifts obtained with the gaussian and the LM2M2 potentials are $-56$ and $-63$ degrees,
respectively. 

As also shown in \cite{kie11}, this anomalous behavior of the soft-core potential at the three-body
level can be corrected by using a short-range effective three-body force, depending only on the hyperradius, 
that is added to the effective potential (\ref{eq4}). In particular we choose the simple three-body
force
\begin{equation}
W(\rho) = W_0 e^{-\rho^2/\rho_0^2},
\label{3b}
\end{equation}
where the strength $W_0$ is adjusted to reproduce the trimer ground-state binding energy 
obtained with the LM2M2 potential. When this is done the results are fairly independent of the
range parameter $\rho_0$, at least within a reasonable value from $\rho_0=4$ a.u. to $\rho_0=10$ a.u..
The results given in the third column (lower part) of Table~\ref{tab3} have been obtained after inclusion 
of the three-body force. The precise values of $W_0$ and $\rho_0$ used are given in the first row
of the lower part of the table.  As we can see, once the binding energy of the ground state has
been corrected, the binding energy of the excited state and the atom-dimer scattering length 
automatically agree with the corresponding values obtained with the LM2M2 potential. Furthermore,
as seen in Fig.3 of \cite{kie11}, the low energy phase-shifts are also corrected when the
effective three-body force is included.

\begin{table}
\caption{Inelasticity ($|{\cal S}_{11}|$) and real part of the phase-shift (Re($\delta$)) for the
$^4$He-$\mbox{$^4$He}_2$ collision at three-body energies (above threshold) $E=5$ mK and
$E=25$ mK. The value of $K_{\mbox{\scriptsize max}}$ is the $K$-value associated
to the last adiabatic potential included in the calculation. The numbers within parenthesis
have been obtained without inclusion of the three-body force.}
\label{tab4}
\begin{tabular}{c|cc|cc}
$K_{\mbox{\scriptsize max}}$   &  \multicolumn{2}{c|}{$|{\cal S}_{11}|$} & \multicolumn{2}{c}{Re($\delta$)} \\ \hline
  & $E=5$ mK  & $E=25$ mK & $E=5$ mK  & $E=25$ mK \\ \hline
  4  & {\bf 0.9988}            &  {\bf 0.9351}            &  {\bf 69.30}            &  {\bf 35.09}  \\
     & {\scriptsize (0.9946)}  &  {\scriptsize (0.9645)}  &  {\scriptsize (75.52)}  &  {\scriptsize (40.59)} \\
  8  & {\bf 0.9988}         &  {\bf 0.9111}               &  {\bf 69.23}  &  {\bf 34.80}  \\
     & {\scriptsize (0.9946)}  &  {\scriptsize (0.9403)}  &  {\scriptsize (75.45)}  &  {\scriptsize (40.28)} \\
 12  & {\bf 0.9989}         &  {\bf 0.9104}               &  {\bf 69.20}  &  {\bf 34.63} \\
     & {\scriptsize (0.9947)}  &  {\scriptsize (0.9394)}  &  {\scriptsize (75.41)}  &  {\scriptsize (40.09)} \\
 16  & {\bf 0.9989}         &  {\bf 0.9110}               &  {\bf 69.17}  &  {\bf 34.60}  \\
     & {\scriptsize (0.9947)}  &  {\scriptsize (0.9402)}  &  {\scriptsize (75.39)}  &  {\scriptsize (40.05)} \\
 20  & {\bf 0.9989}         &  {\bf 0.9110}               &  {\bf 69.16}  &  {\bf 34.59}  \\
     & {\scriptsize (0.9947)}  &  {\scriptsize (0.9402)}  &  {\scriptsize (75.38)}  &  {\scriptsize (40.04)} \\
 24  & {\bf 0.9989}         &  {\bf 0.9109}               &  {\bf 69.15}  &  {\bf 34.58}  \\
     & {\scriptsize (0.9947)}  &  {\scriptsize (0.9402)}  &  {\scriptsize (75.38)}  &  {\scriptsize (40.04)} \\
 28  & {\bf 0.9989}         &  {\bf 0.9109}               &  {\bf 69.15}  &  {\bf 34.58}  \\
     & {\scriptsize (0.9947)}  &  {\scriptsize (0.9402)}  &  {\scriptsize (75.37)}  &  {\scriptsize (40.04)} \\
 40  & {\bf 0.9989}         &  {\bf 0.9109}               &  {\bf 69.15}  &  {\bf 34.58}  \\
     & {\scriptsize (0.9947)}  &  {\scriptsize (0.9402)}  &  {\scriptsize (75.37)}  &  {\scriptsize (40.04)} \\
\end{tabular}
\end{table}

The importance of the inclusion of the three-body force is also seen 
when investigating the $\mbox{$^4$He}-\mbox{$^4$He}_2$ reaction for incident energies ($E_i=E-E_{2b}$) 
above the threshold for breakup of the dimer, i.e., $E_i>|E_{2b}|$ (or $E>0$). In Table~\ref{tab4} we give 
the inelasticity ($|{\cal S}_{11}|$)
and the real part of the phase-shift (Re($\delta$)) for energies $E=5$ and $E=25$ mK.
As in Tables~\ref{tab1} and \ref{tab2}, $K_{\mbox{\scriptsize max}}$ is the 
grand-angular quantum number associated to the last adiabatic term included in the expansion (\ref{eq1}).
The boldface numbers have been obtained when the three-body force has been included in the calculation
($W_0=18.314$ K and $\rho_0=6$ a.u.), and the results within parenthesis have been obtained
without the three-body force.

As we can see, the pattern of convergence is similar to the one observed in Tables~\ref{tab1} and \ref{tab2}
for the neutron-deuteron reaction. A $K_{\mbox{\scriptsize max}}$ value of around 12 is enough to get 
a rather well converged inelasticity, while Re($\delta$) requires a few more adiabatic terms in order to reach
convergence. Similarly to what found in \cite{kie11} for energies below the breakup threshold, 
the inclusion of the three-body force gives rise to relevant changes in the computed values. These changes are 
particularly noticeable for Re($\delta$), which for the two energies under consideration increases up to 
6 degrees when the three-body force is not included in the calculation. Also, we observe that for $E=5$ mK the 
breakup probability ($1-|{\cal S}_{11}|^2$) is still rather small, clearly smaller than 1\% (with three-body
force), while for $E=25$ mK this probability rises up to 17\% (12\% without three-body force).

With the ${\cal S}$-matrix in hand, we can now compute the dissociation rate for the 
$\mbox{$^4$He}-\mbox{$^4$He}_2$ collision. The analytic form of this rate is given in Ref.\cite{sun08}, 
and for $0^+$ states it becomes:
\begin{equation}
D_3= \frac{\hbar \pi}{\mu_{1,23} k_{1,23}} \left( 1- |{\cal S}_{11}|^2\right) ,
\label{disso}
\end{equation}
where
\begin{equation}
\mu_{1,23}=\frac{2M_{\mbox{\scriptsize He}}}{3}\; ,  \hspace*{5mm} 
k_{1,23}^2=\frac{2\mu_{1,23} E_i}{\hbar^2},
\end{equation}
$M_{\mbox{\scriptsize He}}$ is the mass of the $^4$He atom, and $E_i=E-E_{2b}$ is the incident energy 
in the center of mass frame ($E_{2b}$ is the binding energy of the $\mbox{$^4$He}_2$ dimer).

\begin{figure}
\epsfig{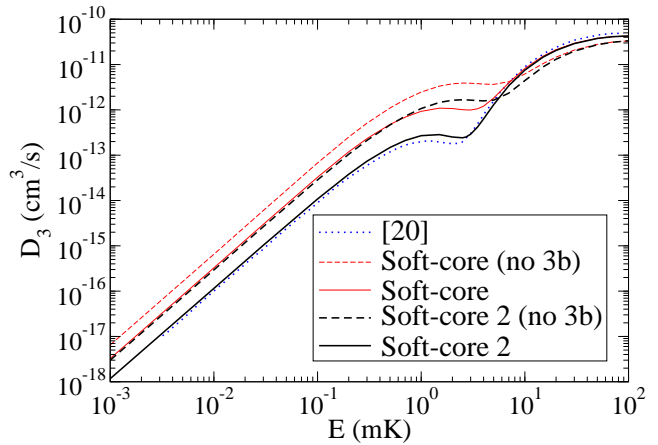}
\caption{(Color online) Dissociation rate (Eq.(\ref{disso})) for the  
$\mbox{$^4$He}+\mbox{$^4$He}_2 \rightarrow \mbox{$^4$He}+\mbox{$^4$He}+\mbox{$^4$He}$ reaction
as a function of the three-body energy $E$. The thin and thick curves have been obtained with the soft-core
potential (\ref{soft}) and (\ref{soft2}), respectively (see Table~\ref{tab3}).  
The dashed and solid curves are the results without and with inclusion of an effective three-body force.
The dotted curve is the result given in \cite{sun08}.  }
\label{fig3}
\end{figure}

In Fig.\ref{fig3} we show the dissociation rate as a function of the three-body energy $E$. The thin-dashed
and thin-solid lines are the results obtained with the gaussian soft-core potential (\ref{soft}) without
and with the additional three-body force (\ref{3b}), respectively.  
The low energy behavior of the dissociation rate follows the 
$E^{K_{\mbox{\scriptsize m}}+2}$ rule derived in \cite{esr01}, where $K_{\mbox{\scriptsize m}}$ is the 
smallest grand-angular quantum number associated to the continuum adiabatic channels ($K_{\mbox{\scriptsize m}}=0$
in our case of three indistinguishable bosons coupled to $J^\pi=0^+$).

From the figure we can see that the effect of the three-body force is quite important. In fact,
for small energies the three-body force reduces the rate by a factor of 2. At higher energies, beyond
10 mK, the effect is the opposite, and the three-body force increases the rate by a factor close to 1.5.

In Ref.\cite{sun08} the same dissociation rate has been computed. The corresponding curve is shown in 
Fig.\ref{fig3} by the dotted curve. As we can see, there is an important difference compared to our 
calculation. Except at high energies ($E \gtrsim 10$ mK), where our result (with three-body force) and the one
in \cite{sun08} basically coincide, for small energies our rate is about a factor of 3 bigger.

However, we have to note that the hard-core two-body potential used in \cite{sun08} gives rise to
somewhat different dimer properties compared to the soft-core potential (\ref{soft}) and the 
LM2M2 potential \cite{azi91}. In \cite{sun08} they have used the potential based on the 
symmetry-adapted perturbation theory (SAPT) derived in \cite{jez07}. The two-body properties obtained with
this potential are given in the last column (upper part) of Table~\ref{tab3}. The two-body dimer is
20\% more bound than with the LM2M2 potential, and therefore the scattering length and the interatomic
distance are smaller than in the LM2M2 case.

To investigate the sensitivity of the dissociation rate to the details of the two-body interaction 
we have then constructed a second gaussian soft-core potential reproducing the two-body properties
of the SAPT potential. This potential takes the form:
\begin{equation}
V_{2b}(r)=-1.234 e^{-r^2/10.03^2},
\label{soft2}
\end{equation}
where the strength is given in K and the range in a.u..
The corresponding two-body properties are given in the fourth column (upper part) of Table~\ref{tab3}
under the label ``soft-core 2''. Again, when moving to the three-body system, this new soft-core potential
presents the same deficiencies as the previous one, i.e., overbinding of the three-body bound states and a too
small atom-dimer scattering length. As before, this problem is solved by inclusion of the three-body force,
whose parameters are again fitted to reproduce the ground state binding energy of the helium trimer provided
by the SAPT potential. The three-body properties with and without three-body force, as well as the 
parameters used for the three-body force are given in the last three columns (lower part) of Table~\ref{tab3}.

Making use of this new soft-core potential and the corresponding three-body force, we can compute again the
dissociation rate (\ref{disso}). The results are shown in Fig.\ref{fig3} by the thick-dashed 
curve (no three-body force) and the thick-solid curve (with three-body force). The new rates are now 
smaller than the ones
obtained with the previous soft-core potential. Furthermore, when the three-body force is included, and 
therefore not only the two-body properties but also the three-body ones agree with the ones obtained
with the SAPT potential, our dissociation rate and the one given in \cite{sun08} agree well for
the whole range of energies. Only a small difference can be seen at about 1 mK. We can therefore see that
the dissociation rate is very sensitive to the details of the two-body interaction. An additional 
20\% binding of the dimer, and the corresponding decrease in the scattering length, leads to a 
factor of 3 decrease in the rate. 

\begin{figure}
\epsfig{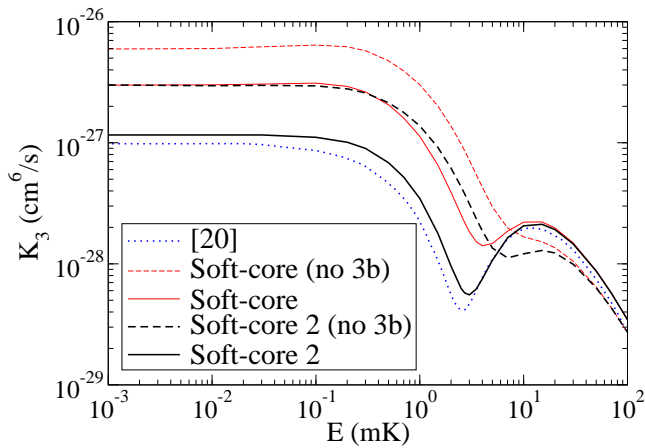}
\caption{(Color online) The same as Fig.\ref{fig3} for the recombination rate (Eq.(\ref{recom})) for the 
process $\mbox{$^4$He}+\mbox{$^4$He}+\mbox{$^4$He}\rightarrow \mbox{$^4$He}_2+\mbox{$^4$He}$.}
\label{fig4}
\end{figure}

The same conclusions are reached when investigating the recombination rate corresponding to the 
inverse process 
$\mbox{$^4$He}+\mbox{$^4$He}+\mbox{$^4$He}\rightarrow \mbox{$^4$He}_2+\mbox{$^4$He}$.
This rate is given by the same ${\cal S}$-matrix elements as in the dissociation case, and its analytical
form for three identical bosons with angular momentum and parity $0^+$ is given by \cite{sun08}:
\begin{equation}
K_3=3! \frac{32\hbar \pi^2}{\mu k^4} \left( 1 - |{\cal S}_{11}|^2\right),
\label{recom}
\end{equation}
where
\begin{equation}
\mu^2=\frac{ M_{\mbox{\scriptsize He}}^2 }{3}
\; , \hspace*{5mm} k^2=\frac{2\mu E}{\hbar^2},
\end{equation}
$M_{\mbox{\scriptsize He}}$ is the mass of the $^4$He atom, and 
$E$ is the total three-body energy.

The computed recombination rate is shown in Fig.\ref{fig4}. The meaning of the curves is as in Fig.\ref{fig3}.
In Ref.\cite{esr01} the recombination rate $K_3$ was proved to behave at low energies as 
$E^{K_{\mbox{\scriptsize m}}}$ ($K_{\mbox{\scriptsize m}}$ is the smallest grand-angular quantum number associated to 
the continuum adiabatic channels), which means that in our case, where $K_{\mbox{\scriptsize m}}=0$,
  $K_3$ should be constant at very low energies, as observed in the figure.  
Again, inclusion of the three-body force reduces the rate by a factor of 2 to 3 at small energies
(compare the dashed curves and the corresponding solid curves in the figure). Also, when the soft-core
potential (\ref{soft2}) together with the three-body force is used, the computed recombination rate agrees
well with the rate given in \cite{sun08} (dotted curve in the figure). This fact reveals again the importance
of the fine details of the two-body interaction.

Finally, it is important to emphasize that in this work the ${\cal S}$-matrix has been obtained by use of
the integral relations (\ref{eq16}) and (\ref{eq17}). As explained, they have the great advantage of 
needing only the 
internal part of the wave functions. In fact, the integrals involved in the calculations shown in this 
section can be safely performed with a maximum value for the hyperradius $\rho$ of 4000 a.u.. Even less 
could be enough. However, in \cite{sun08}, where the adiabatic expansion was also used, the radial wave 
functions in Eq.(\ref{eq1}) had to be expanded up to $5 \times 10^5$ a.u., which is definitely very 
delicate from the numerical point of view.

\section{Summary and conclusions}

In this paper we have extended the use of the integral relations derived in \cite{bar09,rom11} to describe
1+2 reactions above the threshold for breakup of the dimer. As in \cite{bar09,rom11}, the integral relations
are used in combination with the adiabatic expansion method in order to construct the trial wave function.
The main consequence of moving up to energies above the breakup threshold is that the ${\cal S}$-matrix describing
the process has now infinite dimension. This was not the case for energies below the threshold, where the 
elastic, inelastic, or transfer channels were described by a finite (and small) number of adiabatic
terms.

The applicability of the method has been tested with the neutron-deuteron reaction, for which benchmark
calculations are available. The agreement with these calculations is good, and the pattern of convergence
is similar to the one found in \cite{bar09,rom11} for energies below the breakup threshold.  
The integral relations accelerate the convergence significantly, and typically about 10 adiabatic terms
are enough to get a rather well converged result.

The method has then been used to investigate reactions involving three helium atoms. In particular
we have focused on the possibility of using soft-core atom-atom potentials in order to describe
the full process. These potentials permit to avoid all the technical problems arising from the 
use of more sophisticated potentials where a hard-core repulsion is always present.
However, as already shown in \cite{kie11}, even if the soft-core potentials reproduce properly the
dimer properties, an effective three-body force is needed in order to reproduce as well the properties
of the trimer bound states and the atom-dimer scattering length obtained with the hard-core potentials. 

We have found that the three-body force also modifies significantly the inelasticity and the phase-shift
for energies above the breakup threshold. In fact, when computing the reaction rates for dissociation 
of the dimer 
($\mbox{$^4$He}+\mbox{$^4$He}_2 \rightarrow \mbox{$^4$He}+\mbox{$^4$He}+\mbox{$^4$He}$)
and for the recombination process
($\mbox{$^4$He}+\mbox{$^4$He}+\mbox{$^4$He}\rightarrow \mbox{$^4$He}_2+\mbox{$^4$He}$)
the three-body force decreases the rates even by a factor of 3 at low energies.

These two rates are also very sensitive to the details of the two-body interaction. We have found that
a two-body potential, like the SAPT potential, providing a dimer state about 20\% more bound than
the one obtained with the LM2M2 potential, also reduces the rates by a factor of around 3 at small 
energies. The soft-core potential reproducing the two-body properties of the SAPT potential, together
with the corresponding three-body forced designed to reproduce the trimer properties as well, gives then
rise to reaction rates in very good agreement with the ones of the SAPT potential.

In summary, the integral relations are also useful in order to describe reactions above the breakup threshold. 
They accelerate significantly the convergence of the ${\cal S}$-matrix terms. For reactions involving hard-core 
two-body potentials, the use of soft-core potentials with the same two-body properties are a very good 
alternative, provided that they are used together with an effective three-body force designed to fit as well
the bound state three-body energies. When this is done, the reaction rates obtained with the hard-core and
the soft-core potentials agree pretty well. These rates are very sensitive to the details of the two-body
interaction. Small variations of the dimer properties can produce sizable changes in the reaction
rates at low energies. 

\acknowledgments
This work was partly supported by funds provided by DGI of MINECO (Spain) under contract No. FIS2011-23565.

\end{document}